\documentclass[conference]{IEEEtran}
\IEEEoverridecommandlockouts

\usepackage{amsmath,amssymb,amsfonts}
\usepackage{algorithmic}
\usepackage{graphicx}
\usepackage{textcomp}
\usepackage{xcolor}
\usepackage[noadjust]{cite}
\usepackage{braket}
\usepackage{array}
\usepackage{mathtools}
\usepackage{qcircuit}

\begin{document}

\title{Experimental evaluation of quantum Bayesian networks on IBM QX hardware}

\makeatletter
\newcommand{\linebreakand}{%
  \end{@IEEEauthorhalign}
  \hfill\mbox{}\par
  \mbox{}\hfill\begin{@IEEEauthorhalign}
}
\makeatother

\author{\IEEEauthorblockN{Sima E. Borujeni}
\IEEEauthorblockA{\textit{Industrial, Systems, and}\\
\textit{Manufacturing Engineering}\\
\textit{Wichita State University}\\
Wichita, USA \\
sxborujeni@shockers.wichita.edu}
\and
\IEEEauthorblockN{Nam H. Nguyen}
\IEEEauthorblockA{\textit{Boeing Research \& Technology} \\
Huntington Beach, USA \\
nam.h.nguyen5@boeing.com}
\and
\IEEEauthorblockN{Saideep Nannapaneni}
\IEEEauthorblockA{\textit{Industrial, Systems, and}\\
\textit{Manufacturing Engineering}\\
\textit{Wichita State University}\\
Wichita, USA \\
saideep.nannapaneni@wichita.edu}
\linebreakand
\IEEEauthorblockN{Elizabeth C. Behrman}
\IEEEauthorblockA{\textit{Mathematics, Physics, and Statistics} \\
\textit{Wichita State University}\\
Wichita, USA \\
elizabeth.behrman@wichita.edu}
\and
\IEEEauthorblockN{James E. Steck}
\IEEEauthorblockA{\textit{Aerospace Engineering} \\
\textit{Wichita State University}\\
Wichita, USA \\
james.steck@wichita.edu}
}

\maketitle

\begin{abstract}
Bayesian Networks (BN) are probabilistic graphical models that are widely used for uncertainty modeling, stochastic prediction and probabilistic inference. A Quantum Bayesian Network (QBN) is a quantum version of the Bayesian network that utilizes the principles of quantum mechanical systems to improve the computational performance of various analyses. In this paper, we experimentally evaluate the performance of QBN on various IBM QX hardware against Qiskit simulator and classical analysis. We consider a 4-node BN for stock prediction for our experimental evaluation. We construct a quantum circuit to represent the 4-node BN using Qiskit, and run the circuit on nine IBM quantum devices: Yorktown, Vigo, Ourense, Essex, Burlington, London, Rome, Athens and Melbourne. We will also compare the performance of each device across the four levels of optimization performed by the IBM Transpiler when mapping a given quantum circuit to a given device. We use the root mean square percentage error as the metric for performance comparison of various hardware.
 
\end{abstract}


\section{Introduction}
\label{sec:intro}
Quantum computing is a new paradigm of computing that uses the principles of quantum mechanical systems such as superposition and entanglement. This new paradigm has increasing been used to develop algorithms with superior computational performance  when compared to classical counterparts \cite{nielsen2002quantum,kopczyk2018quantum}. The principle of amplitude amplification \cite{brassard2002quantum} has been used to develop computationally efficient algorithms for risk analysis and inference in the context of Bayesian network models \cite{low2014quantum,woerner2019quantum}. Bayesian networks are probabilistic graphical models that are widely used for uncertainty representation and propagation, risk analysis, and probabilistic inference with applications in several domains of science, engineering, and healthcare such as transportation, logistics, bioinformatics, civil infrastructure, manufacturing, and radiotherapy treatment \cite{borujeni2020quantum}. A Quantum Bayesian network (QBN) is a quantum version of the classical Bayesian network. In order to use the developed quantum algorithms, the Bayesian networks should be represented on a quantum computing hardware. 

In our prior work \cite{borujeni2020quantum}, we developed a generic approach to develop a quantum circuit on the IBM quantum gate architecture \cite{dallaire2016quantum} to represent any given Bayesian network. We referred to this approach as Compositional Quantum Bayesian Network (C-QBN) as the overall circuit is obtained by composing smaller circuits relating to marginal/conditional probabilities of various nodes in the Bayesian network. More details are available in Section \ref{sec:qbn}. We demonstrated the developed approach using Qiskit, a Python package from IBM that simulates quantum computing \cite{mckay2018qiskit}.




In addition to the simulator, IBM also provides free access to a number of their real quantum devices with different capacities in terms of the number of qubits \cite{ibmqe}. There are several 5-qubit devices, a 15 qubit device and a simulator available through IBMQ experience, which is an online platform that is used to access the IBM quantum devices. 

As these devices are physical systems, they are affected by several types of noise in the qubits, implementation of quantum gates and in the operating environment. Recently, several studies focus on evaluating the effect of noise, including decoherence, on performance of these devices and finding solutions to mitigate the noise\cite{cross2019validating, harper2019fault, zhang2018efficient}. The computational performance of algorithms that run on the devices also depends on the mapping of variables to various qubits in the devices. Even on the same device, the noise in the implementation of gates on various qubits is different. Since the number of gate operations applied on different qubits is different, it is desired to find the optimal mapping that minimizes the overall error. 

To accomplish this, IBM provides a transpiler that outputs an optimal mapping to qubits on devices. Since different devices have different amounts of gate noise and different connectivity between qubits, the optimal mapping is output considering the gate noise and device connectivity. Assuming all the qubits are connected to each other, the input circuit can be developed using either QASM or Qiskit, and given the choice of device, the transpiler provides a transformed circuit that can be run from that device \cite{Qiskit}. Another input to the transpiler is the amount of optimization that needs to be performed to derive the transformed circuit. There are four levels of optimization, and each level results in a different transformed circuit. More details are available in Section \ref{sec:ibm_qx}.


The goal of this paper is to answer the following questions:
\begin{enumerate}
    \item Can the existing IBM QX hardware simulate Bayesian networks?
    \item Is there a variation in the simulation accuracy across various IBM QX hardware?
    \item Is there a variation in the simulation accuracy across the four levels of optimization available in the transpiler?
\end{enumerate}



\textbf{Paper Organization:} Section. \ref{sec:background} provides a brief background to Bayesian networks and various quantum gates. Section \ref{sec:qbn} discusses the C-QBN approach for the quantum ciruit representation of a QBN. Section. \ref{sec:ibm_qx} discusses various IBM QX hardware, their errors and different levels of the transpiler. Section. \ref{sec:eval} discusses the evaluation study of executing an illustrative 4-node Bayesian network on various hardware and at different levels of optimization followed by concluding remarks and future work in Section \ref{sec:conc}.
    
\section{Background}
\label{sec:background}

\subsection{Bayesian networks}\label{subsec:BN}

A Bayesian Network (BN) is a directed acyclic graphical model with nodes and edges that are used to represent various random variables and dependence among them respectively. Mathematically, a Bayesian network represents a joint probability distribution over a set of random variables as a product of marginal and conditional probability distributions. 


In a BN with $s$ nodes given by set $\mathbb{V}=\{V_1,V_2,...,V_s\}$, the joint distribution can be written as
 
\begin{equation}
P(V_1,V_2,...,V_s)= \prod_{i=1}^{s} P(V_i| \Pi_{V_i})
\end{equation}
where $\Pi_{V_i}$ refers to the set of parent nodes associated with $V_i$. For root nodes (nodes without parent nodes or the nodes at the top of a BN),  $P(V_i| \Pi_{V_i})$ becomes equal to  $P(V_i)$. 

Fig. \ref{fig:2node} represents a simple BN with two discrete variables A and B, each of which can take two values - 0 and 1. Here, A is a root node and the parent node of B. For each value of the parent node (A=0,1), we will have a conditional probability table of the child node (B).

\begin{figure}[!ht]
    \centering
    \includegraphics[scale=0.32]{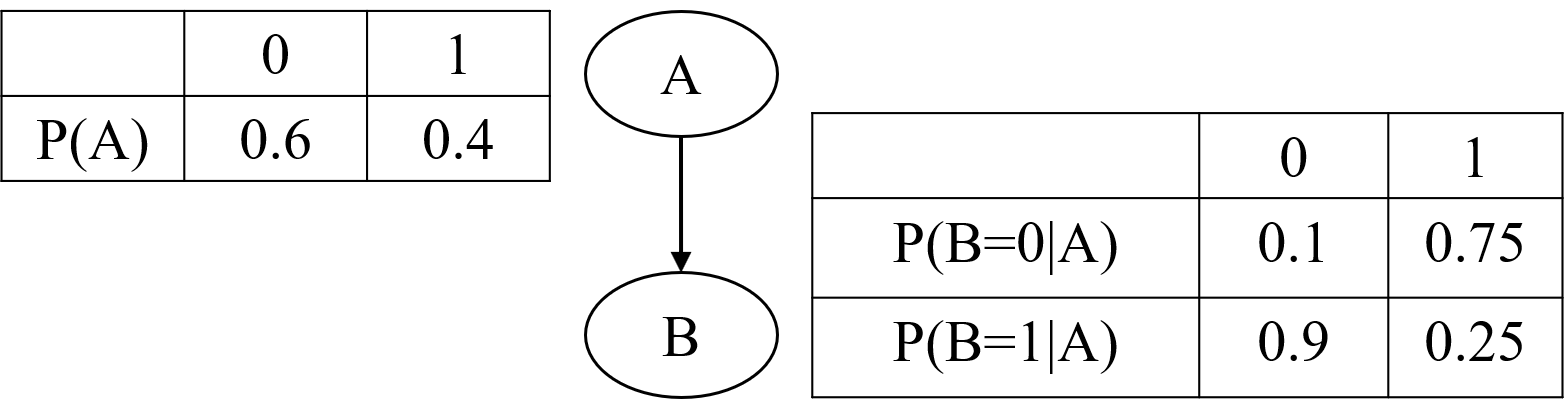}
    \caption{An example of a two-node discrete Bayesian Network}
    \label{fig:2node}
\end{figure}

\subsection{Quantum Gates} \label{subsec:gates}

In this subsection, we will briefly introduce quantum gates that are later used in the development of quantum circuit using the C-QBN approach.

It is a well-known theorem that the set $G=\{ H,T,S,$ CNOT$\}$ formed a universal set of gates for quantum computing \cite{nielsen2002quantum}. Any $n$-qubit unitary operation, represented by a $2^n \times 2^n$ matrix, can be approximated up to an arbitrary precision $\epsilon$ by a sequence of gates consisting of gates from the set $G$. The CNOT (Controlled-NOT) gate is two-qubit gate, where one qubit acts as a control qubit and the other qubit is the target qubit. The matrix representations of various gates using the computational basis of $\Ket{0}$ and $\Ket{1}$ are:

{\renewcommand{\arraystretch}{1.2}
\begin{equation*}\label{HST gates}
       H = \frac{1}{\sqrt{2}}
    \begin{bmatrix} 
    1 & 1\\  
    1 & -1\\ \end{bmatrix}  
    \hspace{0.7 cm}
    S =
    \begin{bmatrix} 
    1 & 0\\  
    0 & i\\ \end{bmatrix}     
    \hspace{0.7 cm}
    T = 
    \begin{bmatrix} 
    1 & 0\\  
    0 & e^{i\pi/4}\\ \end{bmatrix} 
\end{equation*}
}
\begin{equation*}
    \centering
    \text{CNOT}=
    \begin{bmatrix}
I_{2\times2} & 0 \\
0 & X\\
    \end{bmatrix}
\end{equation*}
Here, $I_{2\times2}$ is a $2\times 2$ identity matrix.  The quantum state of the target only changes by a Pauli X gate, when the control is in state $\Ket{1}$ and where
{\renewcommand{\arraystretch}{1.2}
\begin{equation*}\label{PauliX gates}
      X = 
    \begin{bmatrix} 
    0 & 1\\  
    1 & 0\\ \end{bmatrix}  
\end{equation*}}

Rather than being able to implement various single-qubit gates, $H$, $T$ and $S$, IBM hardware can implement an arbitrary single-qubit gate, $U_3(\theta, \phi, \lambda)$ by setting the three parameters, $\theta$, $\phi$ and $\lambda$. Here, $\theta$ represents the angle of rotation about the Y-axis, and $\phi$ and $\lambda$ represent the angles of rotation around the Z-axis in the Bloch sphere \cite{borujeni2020quantum}. The matrix representation of $U_3$ gate is given as

\begin{equation}
\label{eqn:U_3matrix}
 U_3(\theta, \phi, \lambda) = \begin{bmatrix}
\cos\Big(\dfrac{\theta}{2}\Big) &  -e^{i\lambda} \sin\Big(\dfrac{\theta}{2}\Big) \\[2mm]
e^{i\phi} \sin\Big(\dfrac{\theta}{2}\Big) &  e^{i(\phi + \lambda)} \cos\Big(\dfrac{\theta}{2}\Big) \\
\end{bmatrix} 
\end{equation}

However, any multi-qubit gate other than CNOT must be decomposed into a combination of CNOT and single-qubit gates before it can be implemented on the actual hardware.



The generic $U_3$ gate is often called simply $U$ gate. Implementing a general $U_3$ gate can be prone to hardware errors; therefore, IBM allows two additional single-qubit gates, $U_2$ and $U_1$, which are special cases of $U_3$. $U_2 = U(\pi/2, \phi, \lambda) $ and $U_1 = U(0,0,\lambda)$ \cite{ibmqe}. 
We can now represent the single-qubit rotation of $\theta$ around the Y-axis, $R_Y$ gate, as 
\begin{equation}
R_Y(\theta)=U_3(\theta, 0, 0)=
 \begin{bmatrix}
\cos\Big(\dfrac{\theta}{2}\Big) &  -\sin\Big(\dfrac{\theta}{2}\Big) \\[2mm]
\sin\Big(\dfrac{\theta}{2}\Big) & \cos\Big(\dfrac{\theta}{2}\Big)\\
\end{bmatrix}  
\end{equation}

A controlled-U gate, $CU$, is an an application of the gate $U$ to the target qubit(s) when the control qubit is $\Ket{1}$. For example, the controlled-$R_Y$, $CR_Y$, performs the Y-axis rotation of $\theta$ to the target qubit when the controlled qubit is in the state $|1\rangle$. It can be written as, 

\textcolor{black}{\begin{equation}
CR_Y(\theta)=
 \begin{bmatrix}
I_{2\times2} & 0 \\
0 & R_Y(\theta)\\
\end{bmatrix}  
\end{equation}}

The $CU$ gate can be generalized to a general n-qubit controlled gate denoted as $C^nU$ for $n \geq 1$.  Hence in general, $C^nU$ means a unitary operation $U$ will be applied to the target qubit(s) when all the $n$-controlled qubit(s) are in the state $\Ket{1}$. CCNOT (or CCX or Toffoli) and $CCR_Y(\theta)$ are two examples. These are three-qubit gates, with two controlled qubits and one target qubit. The three-qubit gates are not elementary gates; therefore, they are required to be decomposed into a sequence of  single-qubit and CNOT gates \cite{Qiskit}. 





\section{Quantum Bayesian networks}
\label{sec:qbn}

In this section, we discuss the general approach to represent a Bayesian network on the IBM gate architecture using the gates discussed in Section \ref{subsec:gates}, and illustrate the approach for the two-node Bayesian network in Section \ref{subsec:BN}. 

We follow three key ideas when representing a Bayesian network using the gate architecture given below \cite{borujeni2020quantum}.

\begin{enumerate}
    \item Map each node in a BN to one or more qubits (based on the number of discrete states of the random variable)
    \item Map the marginal/conditional probabilities of each node to the probability amplitudes (or probabilities) associated with various states of the qubit(s).
    \item Obtain the desired probability amplitudes of various quantum states through (controlled) rotation gates.
\end{enumerate}  

In C-QBN approach, we represent the marginal/conditional probabilities of each node using appropriate (controlled) rotation gates, and we obtain the overall circuit of the BN by composing the rotation gates of various nodes in the order of the nodes in BN. We start with the root nodes, then represent all child nodes whose parents are the root nodes, and procedure is continued until all the nodes are represented in the quantum circuit. In Fig. \ref{fig:2node}, we begin with the root node (A), and then represent the child node (B).

We begin the circuit with the representation of root nodes using $R_Y$ gates. Applying the rotation gate $R_Y(\theta)$ transforms $\Ket{0}$ to $\cos{\bigg(\dfrac{\theta}{2}\bigg)}\Ket{0} + \sin{\bigg(\dfrac{\theta}{2}\bigg)}\Ket{1}$.
The probabilities associated with $\Ket{0}$ and $\Ket{1}$ states are $\cos^2{\bigg(\dfrac{\theta}{2}\bigg)}$ and $\sin^2{\bigg(\dfrac{\theta}{2}\bigg)}$ respectively. In Fig. \ref{fig:2node}, A is the root node with states 0 and 1. If $P(A=0)$ and $P(A=1)$ represent the probabilities of states 0 and 1, then the rotation angle ($\theta_A$) can be calculated as 

\begin{equation}
\label{eqn:theta}
    \theta_{A} = 2\times\tan^{-1}\sqrt{\dfrac{P(\Ket{1})}{P(\Ket{0})}} = 2\times\tan^{-1}\sqrt{\dfrac{P(A=1)}{P(A=0)}}
\end{equation}

Since the probabilities of a child node are dependent on the values of the parent nodes, we calculate rotation angles for every combination of parent node values, and implement these angles using controlled rotation gates, where the parent nodes act as control qubits and the child node is the target qubit. In Fig. \ref{fig:2node}, B is the child node with A as the parent node. Since A can take two values, we will have two rotation angles of B ($\theta_{B,0}$ and $\theta_{B,1}$) representing its probabilities for $A=0$ and $A=1$ respectively. Fig. \ref{fig:example_cqbn} provides the quantum circuit of the Bayesian network in Fig. \ref{fig:2node}.

\begin{figure}[!ht]
\centering
\scalebox{0.9}{
\hspace*{12mm} \Qcircuit @C=0.2em @R=0.9em {
 && && \mbox{$\ket{0}$}& && & \mbox{$\ket{1}$} & && & & & & &  \\
\lstick{q_0:\Ket{0}}  &\gate{R_Y(\theta_A)}\barrier[0em]{0}&\qw &\gate{X}&  \ctrl{1} &\gate{X}\barrier[0em]{0} &\qw&\qw& \ctrl{1} &\qw&\qw&\qw  \\
\lstick{q_1:\Ket{0}} &\qw \barrier[0em]{0}& \qw &\qw&  \gate{R_Y(\theta_{B,0})} &\qw\barrier[0em]{0}&\qw&\qw& \gate{R_Y(\theta_{B,1})} &\qw&\qw }
}
\vspace{4mm}
\caption{Conceptual quantum circuit associated with the two-node Bayesian network in Fig. \ref{fig:2node}}
\label{fig:example_cqbn}
\end{figure}
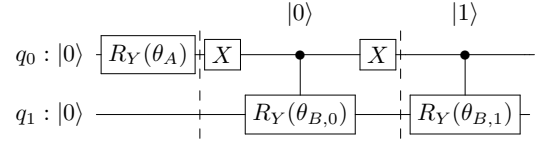

In a $CR_Y(\theta)$ gate, the $R_Y(\theta)$ gate is applied on the target qubit when the control qubit is $\Ket{1}$. Therefore, to represent the conditional probabilities of $B$ when $A=0$, we flip the qubit relating to A using the Pauli X gate (discussed in Section \ref{subsec:gates}), and flip it back after applying the controlled rotation. It should be noted that $CR_Y(\theta)$ is not an elementary gate but Qiskit allows for its application; however, it will be decomposed into a combination of CNOT and $R_Y(\theta)$ gates in the backend.

When the number of parent nodes is greater than 1, then the conditional probabilities of a child node are realized using higher-order controlled rotations ($C^nR_Y(\theta)$).When $n>1$, Qiskit does not allow us to represent $C^nR_Y(\theta)$ gates directly. In order to represent such higher-order rotations, we will use additional qubits called ancilla qubits \cite{borujeni2020quantum}. The example used in the evaluation study (Section \ref{sec:eval}) uses ancilla qubits. Also, more details on representing higher-order controlled rotations are available in \cite{borujeni2020quantum}.


\begin{figure*}[!ht]
    \centering
    \includegraphics[scale=0.62]{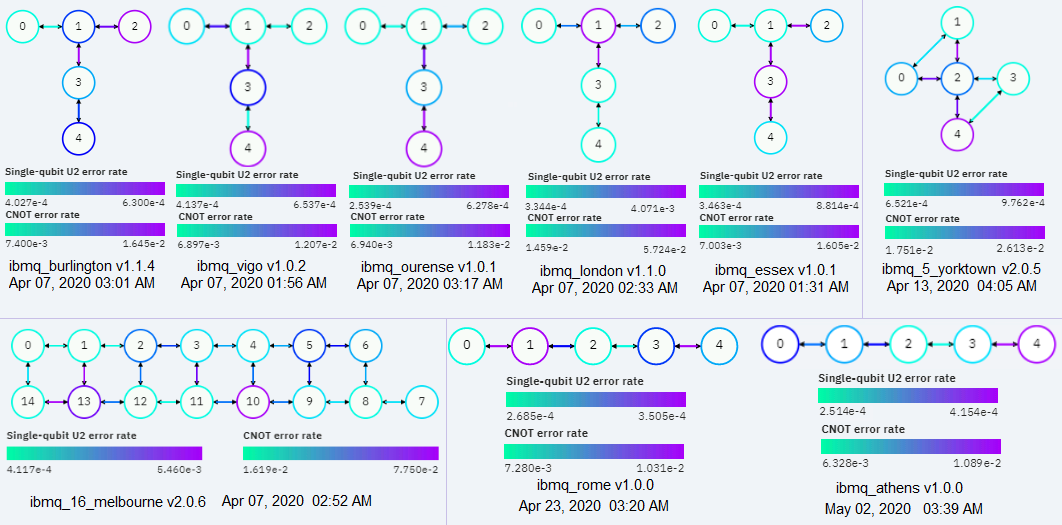}
    \caption{IBM QX hardware devices along with the error rates for single-qubit and CNOT gates at the given calibration time}
    \label{fig:errors}
\end{figure*}

\section{IBM QX Hardware} \label{sec:ibm_qx}

The nine quantum devices that are used in this study are Burlington, Ourense, Vigo, Essex, London, Rome, Athens, Yorktown, and Melbourne. Based on the number of qubits and the architecture, these devices can be divided into four groups. The first group consists of devices with five qubits in a T-shaped architecture. Devices in this group are Burlington, Ourense, Vigo, Essex and London. 
The second group includes devices with five qubits arranged in a line architecture. This group consists of two devices, Rome and Athens. The remaining two devices have two distinct architectures. Yorktown is a five-qubit device with qubits in a bow-tie configuration. Melbourne is a 15-qubit device where the qubits are arranged in a box configuration. In total, we have eight five-qubit devices and one with 15 qubits. The architectures of all the devices are shown in Figure \ref{fig:errors}. The circles represent the qubits and the arrows indicate the connectivity among the qubits.

\textbf{Errors}: Along with the architectures, Fig. \ref{fig:errors} also provides color scales of two types of errors for each device: single-qubit $U_2$ and CNOT error rates. The colors of the qubits (circles) represent the $U_2$ error while the colors of the connections represent the CNOT error. It can be observed that different devices (even with the same architecture) have different single-qubit and CNOT error rates. The $U_2$ error is characterized using randomized benchmarking method \cite{magesan2012characterizing} where a qubit would be taken in a random walk over a route on the Bloch sphere that starts from state $\ket{0}$ and the qubit is expected to go back to $\ket{0}$ in the end. This walk is performed by applying a set of single qubit gates. Increasing the number of these gates will exponentially decrease the chance of going back to the initial state. This decay rate can be used to estimate the average error rate for those single-qubit gates \cite{magesan2012characterizing}. The CNOT error rate is also estimated using a similar approach but using two-qubit Clifford gates \cite{magesan2012characterizing}.

Moreover, the error rates on various devices are periodically updated. Therefore, the Fig. \ref{fig:errors} also provides a timestamp when these error snapshots are taken. From Fig. \ref{fig:errors}, it can noticed that the calibration timestamps for Rome and Athens were more recent when compared against the other devices as they are recently made available by IBM to the public with free access.

\textbf{Transpiler:} As discussed in Section \ref{sec:intro}, the transpiler is used to transpile a given circuit into a circuit that can be executed on a given quantum device after considering the single-qubit and CNOT error rates. Another input to the transpiler for transpiling a given quantum ciruit is the optimization level. The optimization level determines the amount of optimization that needs to be performed in obtaining a transpiled circuit. There are four levels of optimization: \textit{Level 0} (no optimization), \textit{Level 1} (light optimization), \textit{Level 2} (medium optimization), and \textit{Level 3} (heavy optimization). As the optimization level increases, the transpilation time to obtain the optimal implementation of that circuit also increases \cite{Qiskit}.

 
At \textit{Level 0}, there would not be any explicit optimization other than mapping a given circuit to the desired backend device. This is useful for characterization experiments such as randomized benchmarking \cite{magesan2012characterizing} or error amplification where we do not want the transpiler to apply any optimization \cite{Qiskit}. At \textit{Level 1}, the transpiler performs light optimization by collapsing adjacent gates; this combines a chain of single qubit gates (such as rotations) to one gate when feasible. 
\textit{Level 2} optimization includes noise adaptive qubit mapping and gate cancellation by considering commutativity rules for the gates; and finally, \textit{Level 3} optimization includes noise adaptive qubit mapping, gate cancellation using commutativity rules along with unitary synthesis.  
In Qiskit, the default level of optimization is \textit{Level 1} \cite{Qiskit}.


\begin{figure}
    \centering
    \includegraphics[scale=0.26]{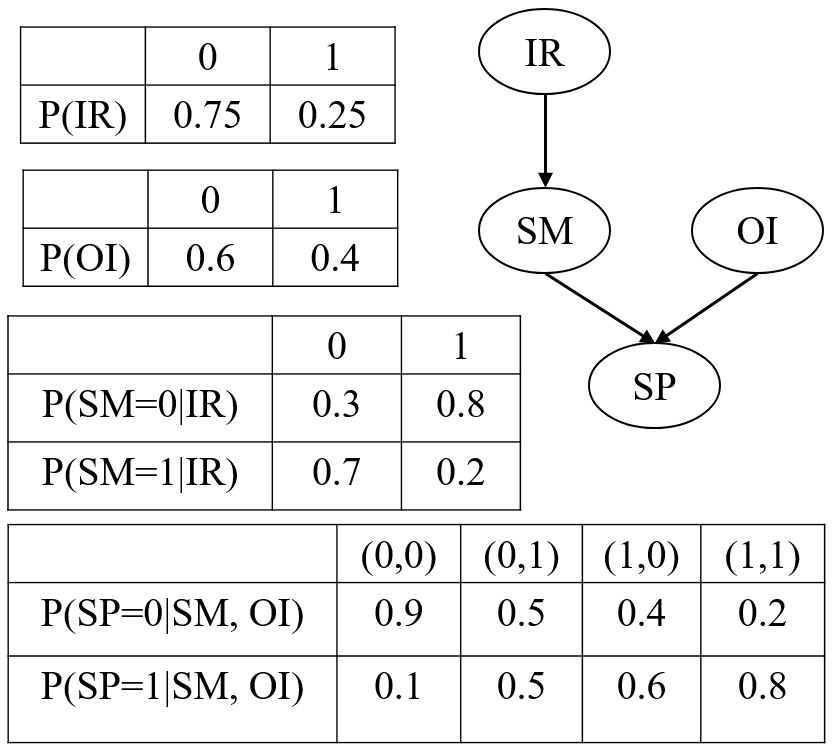}
    \caption{The Bayesian network of the stock prediction for an oil company \cite{shenoy2000bayesian}}
    \label{fig:BNoil}
\end{figure}
    
\section{Evaluation Study}
 \label{sec:eval}  
 
In our evaluation study, we simulated the 4-node Bayesian network given in Fig. \ref{fig:BNoil} on all the nine quantum devices discussed in Section \ref{sec:ibm_qx} along with IBM Qiskit simulator, and compared the results against classical analysis (performed using Netica software \cite{Netica}). The Bayesian network is obtained from \cite{shenoy2000bayesian}, and is used for stock price prediction of an oil company. The variables IR (Interest Rate) and OI (Oil Industry) are the root nodes, and SM (Stock Market) and SP (Stock Price) are the child nodes; these are discrete variables with two values, 0 and 1. In this example, we have two root nodes (IR and OI), and two child nodes, one with one parent node (SM) and the other with two parent nodes (SM, OI).

\textbf{Quantum Circuit:} First, we construct the quantum circuit using the Qiskit package following the procedure discussed in Section \ref{sec:qbn}. This circuit is then run on all the nine hardware devices and the simulator. We considered 8192 shots in each run, as it is the highest number of shots allowed on the IBM devices. The results from each run are used to compute the marginal probabilities of all the nodes. Since each variable takes two values, we will have two marginal probability values, and the sum of them is equal to unity. Therefore, we computed only the probabilities of all the variables being equal to 0 as the probabilities equal to 1 can be obtained by subtracting from unity. Since the measurements obtained from quantum circuits are probabilistic in nature (probabilities based on probability amplitudes of qubits), we performed 10 runs on each device (each run with 8192 shots) and obtained the mean and standard deviation values of the marginal probabilities across the 10 runs. Moreover, we considered all the four optimization levels for each hardware device. This comparison lets us investigate the effect of hardware noise on the accuracy of the results at different optimization levels.

To ensure that the experimental conditions remain constant throughout the runs, all the experiments for each computer were performed simultaneously at the same calibration for all the runs \cite{gottesman2016quantum}. In this way, the variation of noise over time would have the minimal effect on the results. 

\begin{figure*}[!ht]
    \centering
    \includegraphics[scale=0.17]{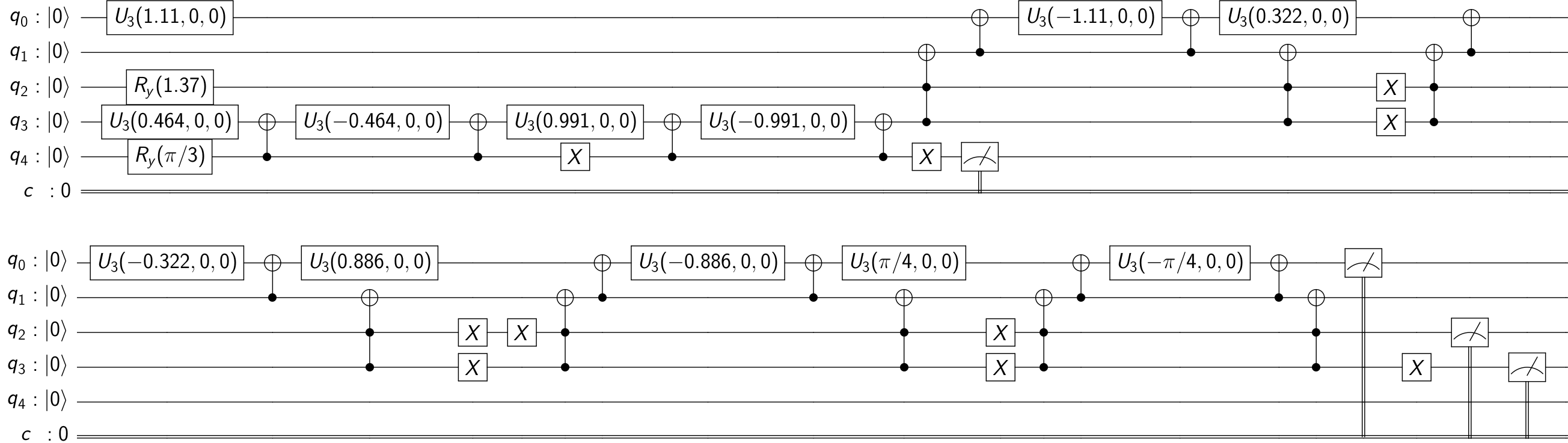}
    \caption{Quantum Circuit of the stock prediction Bayesian network (Fig. \ref{fig:BNoil}) constructed in IBM Qiskit}
    \label{fig:cirq}
\end{figure*}

Fig. \ref{fig:cirq} provides the circuit corresponding to the Bayesian network in Fig. \ref{fig:BNoil}. Since SP has two parent nodes, we will need to implement $CCR_Y(\theta)$ gates to realize its conditional probabilities. As discussed in Section \ref{sec:qbn}, implementation of $CCR_Y(\theta)$ requires the use of an ancilla qubit. Therefore, the quantum circuit in Fig. \ref{fig:cirq} has five qubits, and a measurement bit to store the measurements. In Fig. \ref{fig:cirq}, $q_1$ represents the ancilla and qubits $q_4, q_3, q_2$ and $q_0$ correspond to variables IR, OI, SM, and SP respectively. $CCR_Y(\theta)$ gate is decomposed into a combination of $CR_Y(\theta)$ and CCNOT gates using the ancilla qubit. 

\textbf{Results:} Table \ref{tab:results} provides the mean and standard deviation values of marginal probabilities when run of various IBM hardware at different optimization levels compared with the results from Qiskit simulator and classical analysis. The calibration timestamp and the errors of single-qubit and CNOT gates when the runs are performed are the same as given in Fig. \ref{fig:errors}. Since the simulator is not affected by noise, the results are the same across all the optimization levels.

\begin{table}[htbp]

\caption{Mean and standard deviation values of marginal probabilities over 10 runs on 9 IBM QX hardware compared with the Qiskit simulator and classical analysis (Netica)}
\scalebox{0.9}{
\begin{tabular}{|p{1.25cm}|p{1.52cm}|p{1.52cm}|p{1.52cm}|p{1.52cm}|}
\hline
      & \multicolumn{1}{p{3.2em}|}{\textbf{P(IR=0)}} &  \multicolumn{1}{p{3.2em}|}{\textbf{P(OI=0)}} & \multicolumn{1}{p{3.2em}|}{\textbf{P(SM=0)}} & \multicolumn{1}{p{3em}|}{\textbf{P(SP=0)}} \\
\hline
\textbf{Netica} & 0.750 &       0.600 &       0.425 &        0.499  \\

\textbf{Simulator} & 0.750 (0.006) &       0.601 (0.003)&       0.425 (0.005) &       0.499 (0.006)  \\
\hline
\multicolumn{5}{|c|}{\textbf{Optimization Level 0}} \\
\hline
\textbf{Burlington} & 0.240 (0.015) & 0.536 (0.016) & 0.476 (0.022) & 0.598 (0.022) \\

\textbf{Vigo} & 0.449 (0.021) & 0.544 (0.012) & 0.464 (0.026) & 0.567 (0.018) \\

\textbf{Ourense} & 0.310 (0.028) & 0.529 (0.015) & 0.502 (0.019) & 0.576 (0.017) \\

\textbf{London} & 0.283 (0.021) & 0.558 (0.052) & 0.504 (0.050) & 0.618 (0.064) \\

\textbf{Essex} & 0.330 (0.034) & 0.512 (0.019) & 0.493 (0.022) & 0.562 (0.029) \\

\textbf{Yorktown} & 0.718 (0.024) & 0.567 (0.054) & 0.454 (0.054) & 0.497 (0.025) \\

\textbf{Rome} & 0.393 (0.052) & 0.499 (0.007) & 0.463 (0.007) & 0.674 (0.017) \\

 \textbf{Athens} & 0.410 (0.013) & 0.550 (0.004) & 0.470 (0.011) & 0.601 (0.010)  \\

\textbf{Melbourne} & 0.440 (0.068) & 0.582 (0.024) & 0.506 (0.030) & 0.721 (0.056) \\
\hline
\multicolumn{5}{|c|}{\textbf{Optimization Level 1}} \\
\hline
\textbf{Burlington} & 0.249 (0.069) & 0.523 (0.013) & 0.472 (0.027) & 0.586 (0.024) \\

\textbf{Vigo} & 0.440 (0.035) & 0.547 (0.010) & 0.474 (0.013) & 0.558 (0.013) \\

\textbf{Ourense} & 0.356 (0.032) & 0.515 (0.018) & 0.499 (0.028) & 0.566 (0.020) \\

\textbf{London} & 0.370 (0.021) & 0.540 (0.017) & 0.491 (0.019) & 0.579 (0.021) \\

\textbf{Essex} & 0.332 (0.020) & 0.532 (0.023) & 0.492 (0.018) & 0.578 (0.022) \\

\textbf{Yorktown} & 0.622 (0.158) & 0.579 (0.042) & 0.454 (0.028) & 0.502 (0.026) \\

\textbf{Rome} & 0.386 (0.061) & 0.516 (0.009) & 0.464 (0.007) & 0.669 (0.020) \\

\textbf{Athens} & 0.424 (0.009) & 0.548 (0.008)& 0.463 (0.007) & 0.587 (0.005) \\

\textbf{Melbourne} & 0.449 (0.065) & 0.569 (0.025) & 0.499 (0.070) & 0.709 (0.043) \\
\hline
\multicolumn{5}{|c|}{\textbf{Optimization Level 2}} \\
\hline
\textbf{Burlington} & 0.362 (0.017) & 0.526 (0.022) & 0.461 (0.029) & 0.659 (0.041) \\

\textbf{Vigo} & 0.479 (0.040) & 0.532 (0.022) & 0.476 (0.019) & 0.582 (0.018) \\

\textbf{Ourense} & 0.395 (0.041) & 0.536 (0.023) & 0.491 (0.026) & 0.578 (0.013) \\

\textbf{London} & 0.426 (0.037) & 0.524 (0.012) & 0.509 (0.021) & 0.606 (0.015) \\

\textbf{Essex} & 0.842 (0.015) & 0.551 (0.015) & 0.446 (0.020) & 0.597 (0.023) \\

\textbf{Yorktown} & 0.652  (0.159) & 0.610  (0.015) & 0.462 (0.005) & 0.474 (0.014) \\

\textbf{Rome} & 0.840 (0.017) & 0.541 (0.005) & 0.522 (0.007) & 0.702 (0.014) \\

\textbf{Athens} & 0.832 (0.012) & 0.511 (0.006) & 0.510 (0.006) & 0.630 (0.009) \\

\textbf{Melbourne} & 0.710 (0.185) & 0.574 (0.038) & 0.482 (0.042) & 0.674 (0.049) \\
\hline
\multicolumn{5}{|c|}{\textbf{Optimization Level 3}} \\
\hline
\textbf{Burlington} & 0.526 (0.055) & 0.530 (0.018) & 0.476 (0.031) & 0.620 (0.040) \\

\textbf{Vigo} & 0.525 (0.019) & 0.527 (0.015) & 0.517 (0.020) & 0.561 (0.033) \\

\textbf{Ourense} & 0.497 (0.010) & 0.536 (0.011) & 0.494 (0.020) & 0.567 (0.031) \\

\textbf{London} & 0.552 (0.008) & 0.542 (0.009) & 0.506 (0.019) & 0.601 (0.028) \\

\textbf{Essex} & 0.776 (0.060) & 0.577 (0.026) & 0.483 (0.037) & 0.555 (0.023) \\

\textbf{Yorktown} & 0.626 (0.130) & 0.600 (0.011) & 0.459 (0.014) & 0.465 (0.026) \\

\textbf{Rome} & 0.869 (0.021) & 0.553 (0.006) & 0.503 (0.008) & 0.706 (0.019) \\

\textbf{Athens} & 0.846 (0.020) & 0.526 (0.008) & 0.471 (0.016)  & 0.562 (0.029) \\ 
\textbf{Melbourne} & 0.732 (0.102) & 0.586 (0.025) & 0.474 (0.058) & 0.629 (0.044) \\
\hline
\end{tabular}%
}
\label{tab:results}%
  
\end{table}%

\textbf{Performance comparison:} For comparison of results in Table \ref{tab:results}, we calculated the root mean square percentage error (RMSPE) using the expression in Eq. \ref{eq: epsilon}. The RMSPE error values are given in Table \ref{tab:error_percentage}.

\begin{equation}
\label{eq: epsilon}
\epsilon_T = 100\%\sqrt{\dfrac{1}{n} \sum_{i}  \bigg(\dfrac{p_i^t - \bar{p_i}}{p_i^t}\bigg)^2}
\end{equation}

Here, $\epsilon_T$ is the RMSPE, $p_i^t$ and $\bar{p_i}$ are the true and expectation values (over 10 runs). The true values are obtained from classical analysis, using Netica software.

\begin{figure*}[!ht]
    \centering
    \includegraphics[scale=0.49]{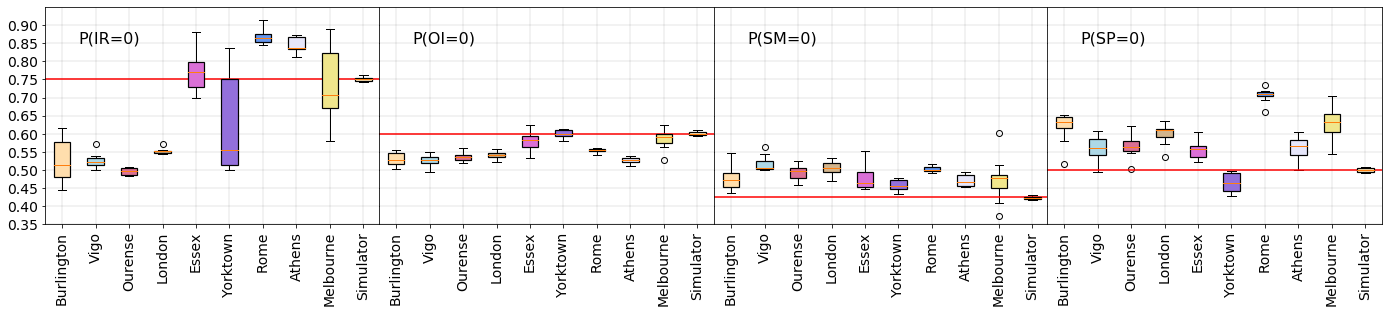}
    \caption{Box plots associated with marginal probability values from all the nine IBM QX hardware at Level 3 optimization, Qiskit, and classical analysis}
    \label{fig:boxpl}
\end{figure*}

\begin{table}[htbp]
  \centering

  \caption{Error rates on various IBM QX hardware at different optimization levels}
    
    \begin{tabular}{|l|c|c|c|c|}
    \hline
     & \multicolumn{1}{p{3.3em}|}{\textbf{ Level 0}} & \multicolumn{1}{p{3.3em}|}{\textbf{ Level 1}} & \multicolumn{1}{p{3.3em}|}{\textbf{ Level 2}} & \multicolumn{1}{p{3.3em}|}{\textbf{ Level 3}} \\
    \hline
    \textbf{Simulator} & 0.1\% & 0.1\% & 0.1\% & 0.1\% \\
    
    \textbf{Burlington} & 36.3\% & 35.6\% & 31.3\% & 21\% \\
    
    \textbf{Vigo} & 11.1\% & 11.4\% & 21.5\% & 20.5\% \\
    
    \textbf{Ourense} & 32.2\% & 29.3\% & 26.7\% & 20.6\% \\
    
    \textbf{London} & 34.8\% & 28.2\% & 26.8\% & 19.8\% \\
    
    \textbf{Essex} & 30.7\% & 30.6\% & 12.5\% & 9.1\% \\
    
    \textbf{Yorktown} & 4.9\% & 9.4\% & 8.3\% & 9.8\% \\
    
    \textbf{Rome} & 31.1\% & 31.8\% & 24.6\% & 24.4\% \\
    
    \textbf{Athens} & 25.8\% & 24.3\% & 18.9\% & 12.2\% \\
    
    \textbf{Melbourne} & 31.9\% & 30.4\% & 19.1\% & 14.4\% \\
    \hline
    \end{tabular}%
  \label{tab:error_percentage}%
\end{table}%

We make the following observations from Table \ref{tab:error_percentage}.
\begin{enumerate}
    \item The results from the simulator are almost the same as the results from classical analysis.
    \item For  most devices (Burlington, Ourense, London, Essex, Rome, Athens, and Melborune), the percentage error decreases with increase in the optimization level.
    \item The error rate for Yorktown is the least at all the optimization levels when compared against all the hardware (at \textit{Level 3}, the error rate for Yorktown was slightly higher than Essex but only by 0.7\%)
    \item The best result across different optimization levels and various hardware is by Yorktown at \textit{Level 0} optimization.
    \item Of all the devices, the error rate significantly decreased for Essex across various optimization levels (30.7\% at \textit{Level 0} to 9.1\% at \textit{Level 3})
\end{enumerate}


Since most devices have the best performance at \textit{Level 3} optimization, we provided box plots in Fig. \ref{fig:boxpl} for all the marginal probabilities to study the variation in results across the 10 runs. In Fig. \ref{fig:boxpl}, the devices are available on the X-axis while probabilities are plotted on the Y-axis. The red line in each plot represents the true probability obtained from classical analysis. For a fair comparison, we used the same scale on Y-axis in all the plots. In Fig. \ref{fig:boxpl}, Yorktown has the largest ranges across all devices; this is particularly evident in the first plot corresponding to P(IR=0).




\section{Conclusion}
\label{sec:conc}

This paper discussed an experimental evaluation of the performance of nine IBM QX hardware (Burlington, Vigo, Ourense, London, Essex, Yorktown, Rome, Athens, and Melbourne) in simulating a Quantum Bayesian network. First, we developed a quantum circuit to represent the Bayesian network using Qiskit, which is Python package for simulating quantum computation. The circuit is then transpiled to run on various hardware, and their performance was compared against that from Qiskit and classical analysis. We also considered all the four levels of optimization (no, light, medium, and heavy optimization) when obtaining a transpiled circuit. On each device, we performed 10 runs, each run with 8192 shots. We used the root mean squared percentage error as a metric to compare the performance of various devices.

We observed that the performance of most devices (6 out of 9) improved with the optimization level. We observed the best performance for Yorktown at all the optimization levels. From the results, we conclude that the existing hardware is not very effective in simulating Quantum Bayesian network due to hardware noise. The error are significant even in a small Bayesian network (with 4 nodes), and we can expect the error rates to increase in more complex Bayesian networks. One way to reduce the error rate could be by developing fault-tolerant circuits and performing fault-tolerant computation.


As our future work, we will consider designing fault tolerant circuits to manage the hardware noise and improve the simulation performance of Quantum Bayesian networks. We will also investigate techniques to develop circuits with lower depths as fewer gates can lead to overall less noisy measurements from the designed circuits.

\bibliographystyle{IEEEtran}
\bibliography{References}
\end{document}